\documentclass[article,aps,pra,onecolumn,groupedaddress,showpacs]{revtex4}
\usepackage{amsmath}
\usepackage{amsfonts}
\usepackage{amssymb}
\usepackage{graphicx}%

\usepackage{color}

\begin{document}

\title[Efficiency of Entanglement Concentration by Photon Subtraction]{Efficiency of Entanglement Concentration 
by Photon Subtraction}

\author{Wulayimu Maimaiti\footnote{ibrahimdulani@gmail.com} and Stefano Mancini }
\address{School of Science and Technology, University of Camerino, I-62032 Camerino, Italy}

\begin{abstract}
We introduce a measure of efficiency for the photon subtraction protocol aimed at entanglement concentration on a single copy of bipartite continuous variable state. 
We then show that iterating the protocol does not lead to higher efficiency than a single application. In order to overcome this limit we present an adaptive version of the protocol able to greatly enhance its efficiency. 
\end{abstract}


\maketitle

\section{Introduction}

The most used states in continuous variable quantum information processing are Gaussian states, such as coherent states, squeezed states, or Einstein-Podolsky-Rosen like states \cite{RMPpirs}. The latter type of entangled states are two mode squeezed vacuum and represents a common source for most 
quantum communication protocols, such as quantum teleportation \cite{telreview}, quantum key distribution
\cite{Reid2000}, quantum dense  coding \cite{1464-4266-1-6-101}, etc.  
However due to the exponential decay of the entanglement 
over the length of quantum communication channels \cite{Welsch2001}, protocols such as entanglement 
concentration are needed to ensure faithful quantum communication \cite{PhysRevLett.76.722}. They deserve to enhance the amount of 
entanglement by using local operations (exploiting ancillary systems) and eventually classical communication.

Entanglement concentration involves several copies of bipartite states each having a low amount of entanglement and aims at producing a smaller 
number of copies possessing a higher amount of entanglement \cite{PhysRevLett.76.722}. However, 
it could also work on a single copy of a bipartite entangled state \cite{PhysRevA.53.2046}, but  
unfortunately it cannot be realized within continuous variable by Gaussian operations \cite{PhysRevLett.89.137903}. 
Hence methods like ``photon subtraction" have been proposed \cite{Kitagawa2006,PhysRevA.82.062316}, where non-Gaussian operations are realized by photon number measurements after
beam splitter transforms with ancillary modes. However the photon subtraction protocol is probabilistic since its 
success depends on the probability of getting `good' measurement outcomes. Hence in characterizing it one should take into account both entanglement enhancement and probability of success. 
To this end we introduce here a measure of efficiency for the photon subtraction protocol and show that iterating it does not lead to higher efficiency than a single application. In order to overcome this limit we present an adaptive version of the protocol able to greatly enhance its efficiency. 
We restrict our analysis to the case of weak fields so to consider only few relevant measurements outcomes. 


\section{Photon Subtraction Protocol}

\begin{figure}
\centering
\includegraphics[scale=0.5]{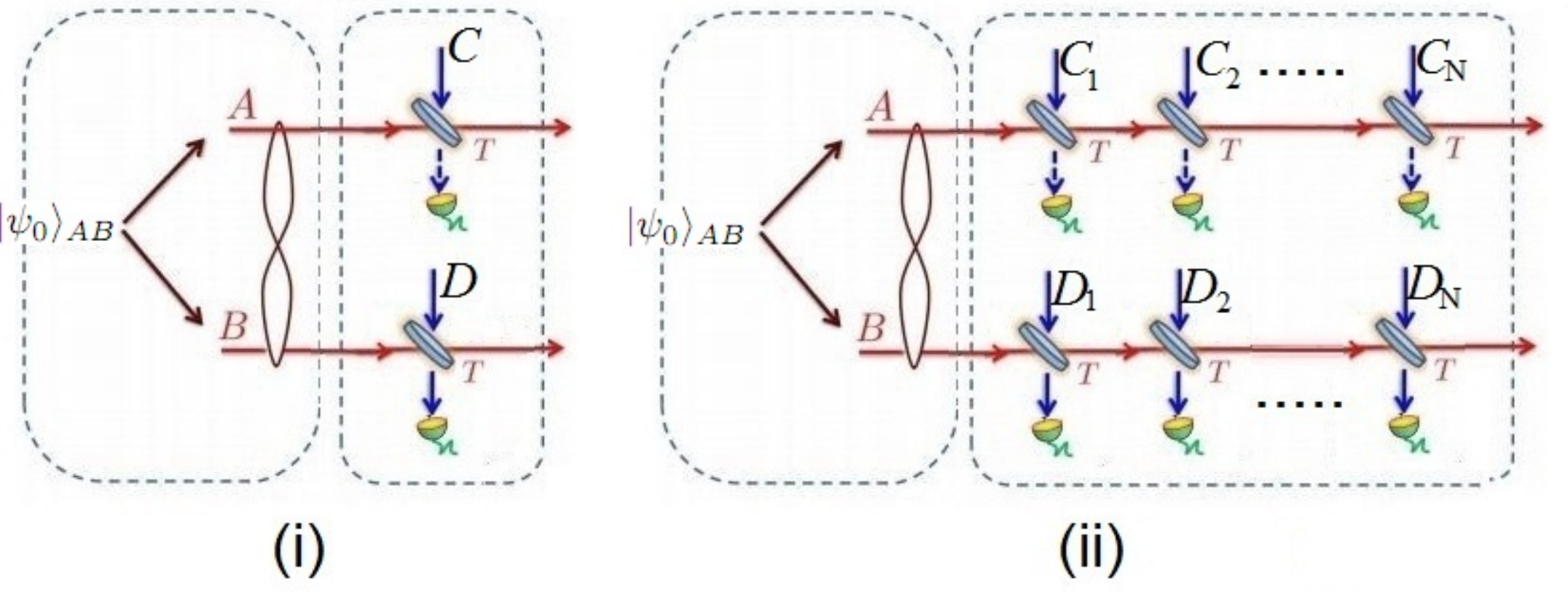}
\caption{Entanglement concentration scheme with photon subtraction. (i) The standard protocol. (ii) The standard protocol iterated $N$ times.}
\label{concentration scheme}
\end{figure}

The standard entanglement concentration scheme with photon subtraction is shown in figure \ref{concentration scheme} (Left). 
The initial state $|\psi_0 \rangle_{AB}$ is a two-mode squeezed vacuum state
\begin{equation}
|\psi_0 \rangle_{AB}
=\sqrt{1-\lambda^2} \sum_{n=0}^{\infty} \lambda^n \vert n \rangle_A \vert n \rangle_B, \quad \lambda \in \mathbb{R}_+.
\label{r2}
\end{equation}
The non-Gaussian operation in photon subtraction scheme is induced by photon number measurement on ancillary modes $C$ and $D$ that interact  with entangled modes $A$ and $B$ through beam splitters of transmittance $T$. 
In Fock space the effect of the beam splitter unitary transformation is \cite{PhysRevA.82.062316}
\begin{equation}
\vert n \rangle_A \vert 0 \rangle_C\stackrel{\hat{U}_{AC}}{\longmapsto}\sum_{k=0}^n
\xi_{nk} \vert n-k \rangle_A \vert k \rangle_C,
\label{BS}
\end{equation}
where
\begin{equation}
\xi_{nk}=(-1)^k \sqrt{
\footnotesize{\left(\begin{array}{c}n\\ k \end{array}\right)}
} (T)^{(n-k)/2} (1-T)^{k/2},
\label{xi}
\end{equation}
with $\footnotesize{\left(\begin{array}{c}n\\ k \end{array}\right)}$ the binomial coefficient. 
Analogous considerations hold true for $BD$ modes.
Now consider the photon number detection of ancillary modes $C,D$.
Suppose the number of photons detected are $c$,$d$ respectively, 
then the corresponding probability operator value elements \cite{Peres1994} are
\begin{equation}
 \hat{M}_{cd}=\vert c \rangle_C \langle c \vert \otimes \vert d \rangle_D \langle d \vert.
\end{equation}
As a consequence, the a posteriori state to this measurement reads
\begin{equation}
\frac{1}{\sqrt{P_{cd}}}  \hat{M}_{cd} \hat{U}_{AC} \hat{U}_{BD} \vert \psi_0 \rangle_{AB} 
 \vert 0 \rangle_C \vert 0 \rangle_D,
\label{basic_scheme_output_state_0}
\end{equation}
where $P_{cd}$ is the joint probability of detecting $c$ (resp. $d$) photons in $C$ (resp. $D$) mode
\begin{equation}
 P_{cd}=\sum_{n=\max[c,d]}^{\infty} (\alpha_n \xi_{nc} \xi_{nd})^2,
\end{equation}
with $\alpha_n=\sqrt{1-\lambda^2}\lambda^n$.
Tracing the state of Eq.(\ref{basic_scheme_output_state_0}) together with its dual over modes $C$ and $D$ we get the following still pure, though 
non-Gaussian, conditional state
\begin{equation}
 \vert \psi_{cd} \rangle_{AB} =\frac{1}{\sqrt{P_{cd}}} \sum_{n=\max[c,d]}^{\infty} \alpha_n \xi_{nc} \xi_{nd} \vert n-c \rangle_A \vert n-d \rangle_B.
\label{basic_scheme_output_state}
\end{equation}


\section{Efficiency of the Protocol}

In order to evaluate the effectiveness of the entanglement concentration protocol we need to compare the prior and posterior amount of entanglement.
A useful entanglement measure is the so called negativity defined as  \cite{PhysRevA.65.032314} 
\begin{equation}
 \mathcal{N}(\rho_{AB}):=\frac{||\rho^{{\sf T}_B}||_1 - 1}{2},
 \label{definition of negativity}
\end{equation}
where $||\hat{O}||_1= {\rm Tr} \sqrt{\hat{O}^{\dagger} \hat{O}}$
is the trace norm of the operator $\hat{O}$ and ${\sf T}_B$ denotes the partial transposition with respect to subsystem 
$B$. 
For pure states with Schmidt decomposition as in Eq.(\ref{basic_scheme_output_state}) the negativity reads
\begin{eqnarray}
\mathcal{N}(\vert \psi_{cd} \rangle_{AB})  =\frac{1}{2} \left(\sum_{n=\max[c,d]}^{\infty} 
\frac{\alpha_n \xi_{nc} \xi_{nd} }{\sqrt{P_{cd}}}
\right)^2 - \frac{1}{2} .\label{negativity of basic output}
\end{eqnarray}
Analogously, the negativity of the initial two mode squeezed vacuum state of Eq.(\ref{r2}) reads
\begin{equation}
 \mathcal{N}_{0} \equiv \mathcal{N}(\vert \psi_0 \rangle_{AB})= \frac{\lambda}{1-\lambda}.
 \label{efficiency def}
\end{equation}
Thus, for a specific measurement outcome $cd$ we can calculate the negativity difference $\mathcal{N}-\mathcal{N}_{0}$ and check if it is positive thus witnessing entanglement enhancement. However this is not enough to evaluate the performance of the scheme. We should also account for the outcome probability. Hence we define the efficiency as   
\begin{equation}
  \mathcal{E}:=\sum_m P_m \frac{\Delta \mathcal{N}(\vert \psi_m \rangle)}{\mathcal{N}(\vert \psi_m \rangle)},
  \label{effdef}
\end{equation}
where the index $m$ runs overall possible measurement outcomes and $\Delta \mathcal{N}(\vert \psi_m \rangle)$ is  defined as
\begin{equation}
\Delta \mathcal{N}(\vert \psi_m \rangle):=\left\{\begin{array}{ccc}
\mathcal{N}(\vert \psi_m \rangle)-\mathcal{N}_0 & \mathrm{if} & \mathcal{N}(\vert \psi_m \rangle)-\mathcal{N}_0>0\\
0 &  \mathrm{if} & \mathcal{N}(\vert \psi_m \rangle)-\mathcal{N}_0\le 0
\end{array}\right. .
\label{delta N definition}
\end{equation}
Notice that only the measurement outcomes with increased negativities contribute 
to the efficiency. Furthermore it is $0\le {\mathcal E}\le 1$.


\section{Iteration of The Protocol}

In this Section we consider the iteration of the standard photon subtraction scheme as shown in Figure \ref{concentration scheme} (Right). This means to have $N$ ancillary modes $C_i$, $i=1,\ldots,N$ (resp. $D_j$, $j=1,\ldots,N$) on arm $A$ (resp. $B$) and couple each of them sequentially with $A$ mode (resp. $B$ mode).

Suppose in the $A$ (resp. B) arm $c_i$ (resp. $d_j$) photons are detected in the $i$-th (resp. $j$-th) ancilla and 
zero everywhere else, then the resultant state can be straightforwardly computed following Eq.(\ref{basic_scheme_output_state})
\begin{eqnarray}
 \vert \psi_{c_id_j} \rangle_{AB} = \frac{1}{\sqrt{P_{c_i,d_j}}} \sum_{n=\max[c_i,d_j]}^{\infty} 
& \alpha_n \xi_{n,0}^{i+j-2} \xi_{n,c_i} \xi_{n,d_j} 
\xi_{n-c_i,0}^{N-i}  \xi_{n-d_j,0}^{N-j}\nonumber \\ &\times \vert n-c_i \rangle_A \vert n-d_j \rangle_B,
\label{the state for k,l photon in i,j ancillary modes}
\end{eqnarray}
where $P_{c_i,d_j}$ is the probability of detecting $c_i$ photons in $i$-th ancilla in $A$ arm and $d_j$ photons in $j$-th ancilla in $B$ arm
\begin{equation}
 P_{c_i,d_j}= \sum_{n=\max[c_i,d_j]}^{\infty} {(\alpha_n \xi_{n,0}^{i+j-2} \xi_{n,c_i} \xi_{n,d_j} 
\xi_{n-c_i,0}^{N-i} \xi_{n-d_j,0}^{N-j})}^2.
\label{serial N detection probability}
\end{equation}
Of course we should consider not only the photons detected in the $i$-th ancilla of $A$ arm and $j$-th ancilla of $B$ arm, but on all ancillary modes of the $A$ arm as well as of the $B$ arm.
For the sake of convenience we put some limits in our consideration of the number of detected photons. 
Actually we restrict our attention to $\lambda\in(0,0.325)$ which according to Eq.(\ref{r2}) guarantees that the probability of having two or more photons on each arm is negligible small (smaller than 1\%).
This is also consistent with the fact that entanglement purification is usually required when
the amount of initial entanglement (value of parameter $\lambda$) is quite low.

Therefore from now on we will consider single photon detection in each arm. Then, we can divide the possible outcomes into two classes: \emph{symmetric} having single photon detection in both arms ($\sum_{i=1}^Nc_i=\sum_{j=1}^N d_j=1$) and \emph{asymmetric} having single photon detection only in one arm (either $\sum_{i=1}^Nc_i=0, \sum_{j=1}^N d_j=1$ or $\sum_{i=1}^Nc_i=1, \sum_{j=1}^N d_j=0$).

\subsubsection{Symmetric case}

We can use Eq.(\ref{serial N detection probability}) to evaluate the probability of detecting one photon in the $i$-th ancilla of $A$ arm and one photon in the $j$-th ancilla of $B$ arm. It results
\begin{equation}
P_{1_i,1_j} =\frac{(1-T)^2 T^{-2+i+j}\lambda^2(1-\lambda^2) (1+T^{2N}\lambda^2)}{(1-T^{2N}\lambda^2)^3}.
\label{N BS symmetric dp}
\end{equation}
Clearly the probability of detecting one photon in first ancillary modes ($i=j=1$) is the highest. 
The evaluation of negativity for single photon detected on each arm turns out to be equivalent to the evaluation of negativity for the state of Eq.(\ref{the state for k,l photon in i,j ancillary modes}).
Actually we have
\begin{equation}
 {\mathcal N} (\vert \psi_{1_i,1_j} \rangle_{AB}) = \frac{T^N\lambda (2+T^N\lambda+T^{2N}\lambda^2)}{(1-T^N\lambda)(1+T^{2N}\lambda^2)},
 \label{neg1i1j}
\end{equation}
no matter on which ancilla the photon is detected. Notice that this negativity decreases by increasing $N$.

\subsubsection{Asymmetric case}

Using Eq.(\ref{serial N detection probability}) we can get the probability of detecting one photon in $i$-th ancilla in one arm  
\begin{equation}
P_{1_i, 0} = P_{0,1_i}= \frac{T^{-1+i+N}(1-T)\lambda^2(1-\lambda^2)}{(1-T^{2N}\lambda^2)^2}.
\label{N BS asymmetric dp}
\end{equation}
The negativity of the state (\ref{the state for k,l photon in i,j ancillary modes}) corresponding to a photon detection in $i$-th ancilla in one arm results
\begin{equation}
 {\mathcal N}(\vert \psi_{1_i,0} \rangle_{AB})= \frac{(T^{2N}\lambda^2-1)^2}{2\lambda^2T^{2N}} 
 \left[\mathrm{Li}_{-\frac{1}{2}}\left(\lambda T^N\right) \right]^2 -\frac{1}{2},
\label{serial N asymmetric neg}
\end{equation}
where $\mathrm{Li}_{-\frac{1}{2}}(\bullet)$ is the polylogarithm function of order $-\frac{1}{2}$.
Also in this case the negativity does not depend on the outcome and decreases with $N$.

\subsubsection{Overall efficiency}

According to the definition of Eq.(\ref{effdef}), we write the overall efficiency of iteration protocol as
\begin{equation}
\label{effsum}
{\cal E}=\sum_{i,j=1}^N P_{1_i,1_j}  \frac{\Delta{\mathcal N} (\vert \psi_{1_i,1_j} \rangle_{AB})}{{\mathcal N} (\vert \psi_{1_i,1_j} \rangle_{AB})}
+2\sum_{i=1}^N P_{1_i,0}  \frac{\Delta{\mathcal N} (\vert \psi_{1_i,0} \rangle_{AB}}{{\mathcal N} (\vert \psi_{1_i,0} \rangle_{AB})}.
\end{equation}
where $\Delta{\mathcal N} (\vert \psi_{1_i,1_j} \rangle_{AB} $ and $\Delta{\mathcal N} (\vert \psi_{1_i,0} \rangle_{AB}$ are as defined in Eq.(\ref{delta N definition}). The factor 2 in front of the second term come from the fact that $\vert \psi_{1_i,0} \rangle_{AB}=\vert \psi_{0,1_i} \rangle_{AB}$ and $P_{1_i,0}=P_{0,1_i}$.

\begin{figure}[h]
 \centering
\includegraphics[scale=0.32]{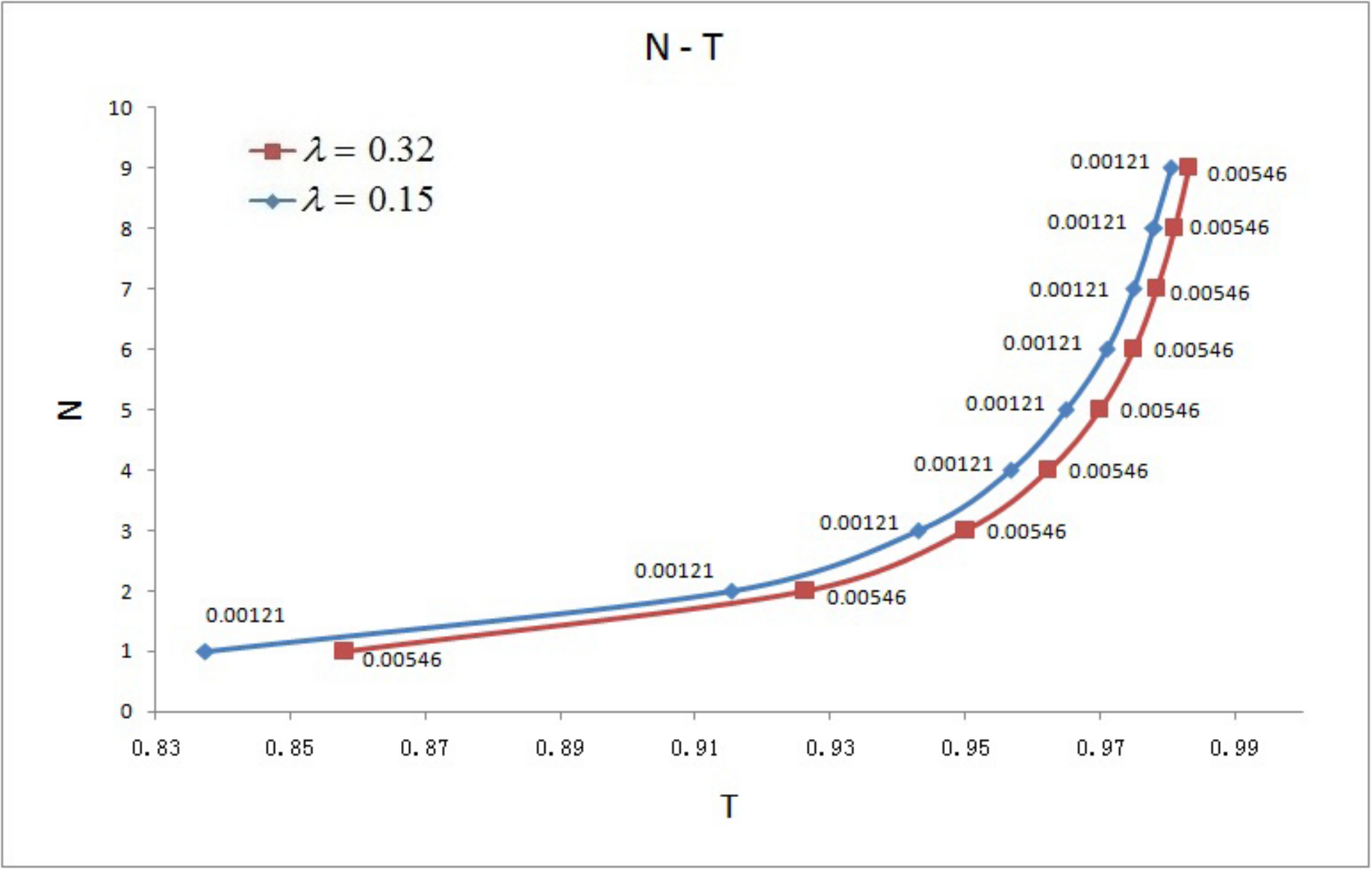}\,\includegraphics[scale=0.32]{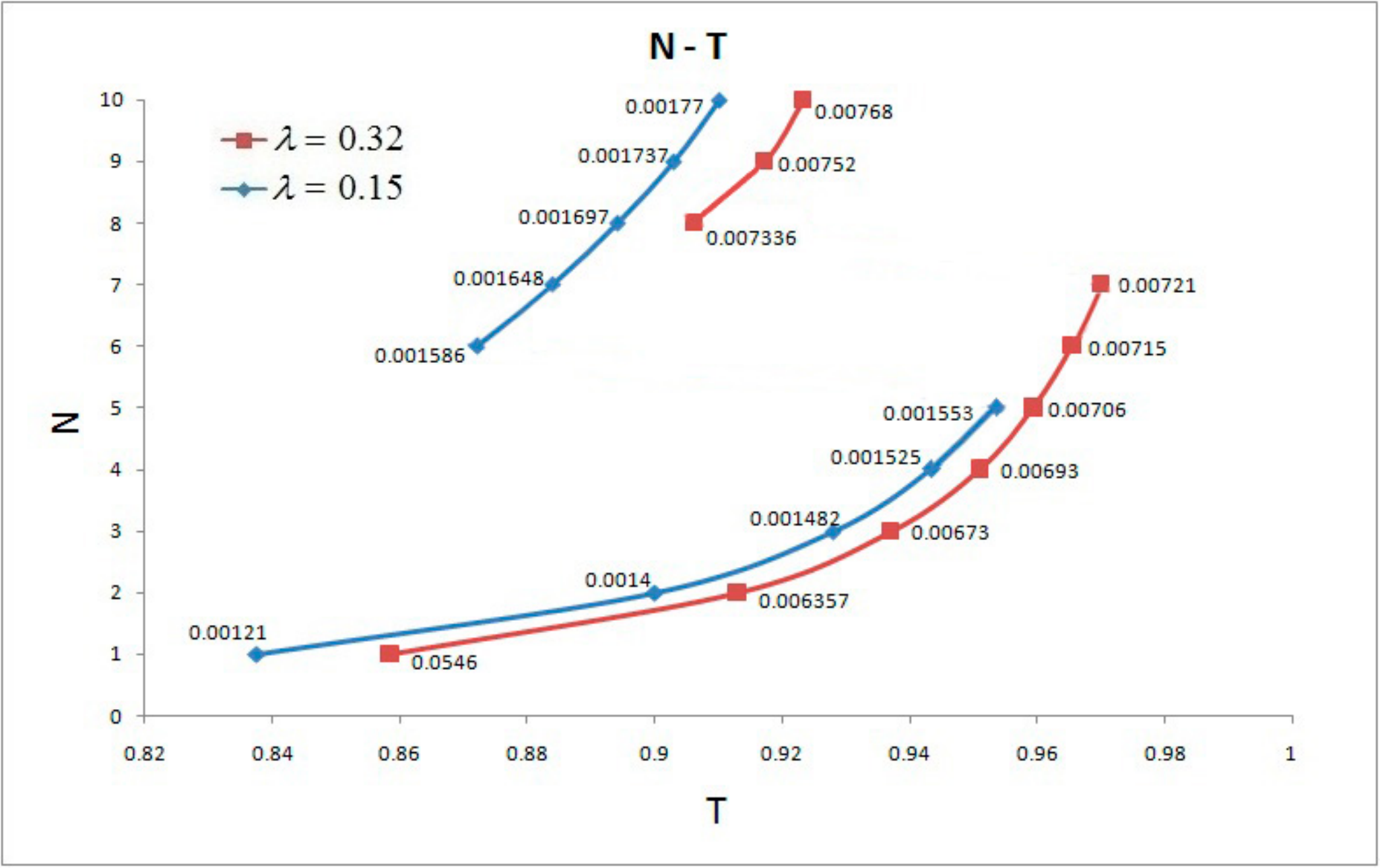}
\caption{Loci of the points in the $N-T$ plane giving maximum efficiency for iteration of the standard protocol (Left) and  adaptive protocol (Right) when $\lambda=0.15$ and $0.32$. The values of efficiency are reported beside the points.}
\label{figures}
\end{figure}

Using Eqs.(\ref{N BS symmetric dp}), (\ref{neg1i1j}) and Eqs.(\ref{N BS asymmetric dp}), (\ref{serial N asymmetric neg}) we get ${\cal E}$ as function of $(\lambda,N,T)$
\begin{equation}
{\cal E}=\frac{(1-T^N)\lambda^2(1-\lambda^2)}{1-\lambda^2 T^{2N}}\left[
\frac{(1-T^N)(1+\lambda^2T^{2N})}{1-\lambda^2 T^{2N}}E_1+2T^NE_2
\right],
\end{equation}
with
\begin{eqnarray}
&E_1=\max\left[0,1-\frac{(1-\lambda T^N)(1+\lambda^2T^{2N})}
{T^N(1-\lambda) (2+\lambda T^N+\lambda^2T^{2N})}
\right],\\
&E_2= \max\left[0, 
 1-\frac{\lambda}{1-\lambda} \frac{2\lambda^2T^{2N}}{(1-\lambda^2T^{2N})^2
 \left[\mathrm{Li}_{-\frac{1}{2}}\left(\lambda T^N\right)
 \right]^2-T^{2N}\lambda^2}\right].
\end{eqnarray}
In Figure \ref{figures} (Left) we show the loci in the $N-T$ plane where ${\cal E}$ reaches its maximum for given values of $\lambda$. Interestingly the maximum values taken by ${\cal E}$ are the same for fixed $\lambda$. This means that for a given value of $\lambda$ the same efficiency can be reached for many pair values of $N-T$ including a pair with $N=1$.  

\section{Adaptive Scheme}

In this section we introduce a possible way to avoid the reduction of negativity increment by $N$ while 
keeping high success probability. The idea is to apply on each arm a beam splitter conditionally to no photon detection in the previous one. Whenever there is a detection no more beam splitters will be applied afterwards. This can be done with the help of some feed-forward control system. 

According to this strategy all the terms related to no photon detection  will be eliminated from the coefficients of the state in Eq.(\ref{the state for k,l photon in i,j ancillary modes}) and from the detection probability in Eq.(\ref{serial N detection probability}) after a photon detection, namely the state in Eq.(\ref{the state for k,l photon in i,j ancillary modes}) becomes
\begin{equation}
 \vert \psi_{c_id_j} \rangle_{AB} = \frac{1}{\sqrt{P_{c_i,d_j}}}  \sum_{n=\max[c_i,d_j]}^{\infty} \alpha_n \xi_{n,0}^{i+j-2} \xi_{n,c_i} \xi_{n,d_j} 
 \vert n-c_i \rangle_A \vert n-d_j \rangle_B,
\label{the state for c,d photon in i,j ancillary modes in adaptive scheme}
\end{equation}
with $ P_{c_i,d_j}$ now reading
\begin{equation}
 P_{c_i,d_j}= \sum_{n=\max[c_i,d_j]}^{\infty} {(\alpha_n \xi_{n,0}^{i+j-2} \xi_{n,c_i} \xi_{n,d_j} )}^2.
\label{serial N detection probability in adaptive scheme}
\end{equation}
Limiting the maximum number of detected photons to 1, we have likewise the previous Section symmetric and asymmetric cases.

\subsubsection{Symmetric case}
Setting $c_i=d_j=1$ in Eqs.(\ref{the state for c,d photon in i,j ancillary modes in adaptive scheme}), (\ref{serial N detection probability in adaptive scheme}) we get
\begin{equation}
P_{1_i,1_j} =\frac{(1-T)^2 T^{-2+i+j}\lambda^2(1-\lambda^2) (1+T^{i+j}\lambda^2)}{(1-T^{i+j}\lambda^2)^3},
\label{Psymada}
\end{equation}
and the negativity of symmetric outcomes 
\begin{equation}
 \mathcal{N}\left(\vert \psi_{1_i,1_j} \rangle_{AB}\right) =   \frac{(1-T^{i+j}\lambda^2)^3}{2(1-T^{(i+j)/2}\lambda)^4
 (1+T^{i+j}\lambda^2)}-\frac{1}{2}.
\label{adaptive re-iteration negativity}
\end{equation}

\subsubsection{Asymmetric case}
In the same way, setting $c_i=1, d_j=0$ in Eqs.(\ref{the state for c,d photon in i,j ancillary modes in adaptive scheme}), (\ref{serial N detection probability in adaptive scheme}) yields 
\begin{equation}
P_{1_i,0} = P_{0,1_i}= \frac{T^{-1+i+N}(1-T)\lambda^2(1-\lambda^2)}{(1-T^{i+N}\lambda^2)^2},
\label{Pasyada}
\end{equation}
and the negativity of asymmetric outcomes 
\begin{equation}
\mathcal{N}\left(\vert \psi_{1_i,0} \rangle_{AB}\right) = \frac{(1-T^{i+N}\lambda^2)^2}{2\lambda^2T^{(i+N)}} 
\left[
\mathrm{Li}_{-\frac{1}{2}}\left(\lambda T^{(i+N)/2}\right)
\right]^2-\frac{1}{2}.
\label{adaptive re-iteration negativity-asymmetric case}
\end{equation}

\subsubsection{Overall efficiency}
To evaluate the overall efficiency of adaptive scheme
we can insert Eqs.(\ref{Psymada}), (\ref{adaptive re-iteration negativity-asymmetric case}) and Eqs.(\ref{Pasyada}), (\ref{adaptive re-iteration negativity-asymmetric case}) into Eq.(\ref{effsum}). However in this case is not possible to get a compact expression for it. Results from numerical evaluation are shown in
Figure \ref{figures} (Right). In this case the maximum values taken by ${\cal E}$ for fixed value of $\lambda$ are not the same, and monotonically increase towards ${\cal E}=1$ for $N\to\infty$ and $T\to 1$ (though not along a unique line in the $N-T$ plane).


\section{Conclusion}

We have considered the efficiency of entanglement concentration by photon subtraction defined as 
the average of entanglement increments weighted by their success probability. While entanglement increment diminishes by iterating the protocol, its success probability auguments. It results that these two behaviors compensate each other so that optimal efficiency can be already reached in one step of the protocol. 

In contrast, iteration of photon subtraction becomes meaningful when no photon have been previously detected. Thus adaptive strategy can in principle enhances the efficiency up to one. Implementing it is challenging because
light should be stored during the feedforward time, however quantum memories \cite{qm} are promising for achieving significant values of efficiency in this way.


\end{document}